# Multimodal Physical Fitness Monitoring (PFM) Framework Based on TimeMAE-PFM in Wearable Scenarios

Junjie Zhang
School of Computer Science and Technology
Sichuan University
Sichuan, China
junjiezhang1@vip.qq.com

Zheming Zhang
School of Computer Science and Technology
Sichuan University
Sichuan, China
zheming.z@foxmail.com

Huachen Xiang
School of Computer Science and Technology
Sichuan University
Sichuan, China
1487543516@qq.com

Yangquan Tan
West China School of Clinical Medicine
Sichuan University
Sichuan, China
1294877931@qq.com

Linnan Huo
West China School of Clinical Medicine
Sichuan University
Sichuan, China
710109533@qq.com

Fengyi Wang
Department of Rehabilitation Medicine
West China Hospital, Sichuan University
Sichuan, China
* Corresponding author: 1187584854@qq.com

*Abstract*—Physical function monitoring (PFM) plays a crucial role in healthcare especially for the elderly. Traditional assessment methods such as the Short Physical Performance Battery (SPPB) have failed to capture the full dynamic characteristics of physical function. Wearable sensors such as smart wristbands offer a promising solution to this issue. However, challenges exist, such as the computational complexity of machine learning methods and inadequate information capture. This paper proposes a multi-modal PFM framework based on an improved TimeMAE, which compresses time-series data into a low-dimensional latent space and integrates a self-enhanced attention module. This framework achieves effective monitoring of physical health, providing a solution for real-time and personalized assessment. The method is validated using the NHATS dataset, and the results demonstrate an accuracy of 70.6% and an AUC of 82.20%, surpassing other state-of-the-art time-series classification models.

*Keywords-Physical PFM; elderly; wearable sensors; TimeMAE; self-enhanced attention module; personalized evaluation*

## I. INTRODUCTION

Physical fitness monitoring (PFM) is an essential aspect of healthcare, particularly for older adults, as it significantly impacts overall health and well-being. Regular PFM plays a crucial role in identifying potential health risks, guiding interventions, and promoting healthy ageing. However, traditional assessment methods face challenges as they are often benchmarked against the Short Physical Performance Battery (SPPB)[1]. The SPPB evaluates different aspects of physical functioning, including balance, gait speed, and lower extremity strength, and provides valuable information for understanding overall health. However, relying solely on intermittent clinic assessments may not fully capture the dynamic nature of physical fitness, potentially overlooking important functional abilities in real-world settings.

Wearable sensors, such as smart bracelets or actigraph, offer a promising solution to the limitations of traditional monitoring methods. These devices enable continuous and unobtrusive monitoring of physical activity in natural environments, allowing for remote assessment and intervention. By providing real-time data on movement patterns, sleep quality, and activity levels, wearable sensors offer valuable insights into an individual's daily habits and behaviours. This information enables personalised interventions and proactive health management. Activity sequences captured by activity recorders (Actidata for short) often include accelerometer data sequences collected over long periods, providing insight into an individual's activity patterns, including periods of rest, moderate activity, and strenuous exercise. Furthermore, Actidata applications frequently enhance these activity records with fundamental demographic details, such as age, gender, and body mass index (BMI)[2]. The inclusion of demographic data enhances the context of activity data, enabling more comprehensive analyses and interpretations.

Although Actidata-based PFM implementations have great potential, there are significant challenges. Machine learning-based approaches typically require manual computation of a large number of features extracted from accelerometer data, which exacerbates the curse of dimensionality and computational complexity. Recurrent neural networks (RNNs) may have difficulty retaining relevant information over long time series, potentially overlooking subtle patterns indicative of physical health. Additionally, effective integration of demographic information with Actidata to reveal meaningful relationships may require specialized techniques to capture the complexity inherent in multimodal data.

To address these challenges, we suggest an enhanced multimodal PFM framework that utilises TimeMAE. The central aspect of the improved TimeMAE is to condense the input time-series data into a low-dimensional potential space to capture the crucial features of the data. Afterwards, the decoder reconstructs the original input using unsupervised auto-coding techniques to minimise the reconstruction error. The encoder learns a condensed representation of the input time series data, reducing its dimensionality while effectively capturing temporal dependencies and patterns in Actidata. Furthermore, our framework includes a self-enhancing attention module that seamlessly integrates demographic information with Actidata to facilitate comprehensive physical health monitoring. The framework combines TimeMAE and TabSRA to offer a robust solution for real-time, personalised physical health assessment in wearable scenarios.

## II. METHODS

The network architecture we proposed based on a partial TimeMAE[3] structure. It includes a temporal self-supervised autoencoder that only contains masked representational regression tasks. Additionally, we designed a multimodal informative attentional self-enhancement method to complete the entire training process. Figure 1 shows the complete network structure. The network is divided into two training phases. In the encoder pre-training phase, input sequences are divided into visible and masked inputs through a masking strategy. These inputs are then projected into latent representations through a convolutional layer. Two positional representations are then learned with two encoders using the decoupled encoder architecture. The visible sequence encoder undergoes self-supervised training to reconstruct the masked sequence. In the full learning phase, we input the extracted temporal features and demographic data. Our designed self-feature augmented attention module obtains a set of multimodal fusion reinforced features, which undergo a linear mapping layer to produce the final classification.

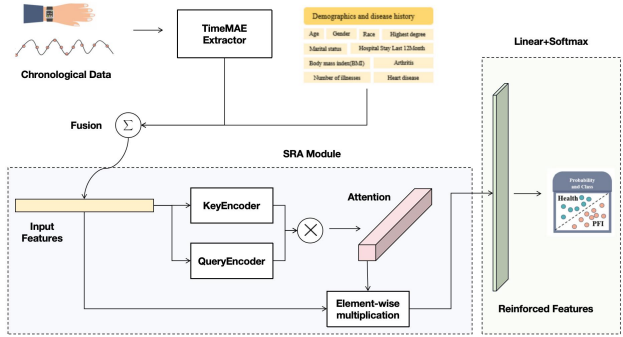

Figure 1. Overall flow chart

### A. Feature Encoder and Random Mask

The current default paradigm for characterising time-series data is point-based, which may miss important semantic features and does not allow e allow the encoder to identify subtle patterns that indicate the active health of the elderly.

To improve information density, it is necessary to enable the encoder to learn motion patterns from local sub-sequences. The concept of TimeMAE is used to define a time sequence $X = x_1, x_2, ..., x_T \in \mathbb{R}^{T \times m}$, with slices defined as $s_{i,j} = x_i, x_{i+1}, ..., x_{i+\sigma}$, where the sub-sequences have a length of $\sigma$. It is important to note that the visible and masked parts are independent, and any insufficient parts of the sequence are zero-padded. The output is represented as $Z = [z_1, z_2, ...z_{\lceil T/\sigma \rceil}] \in \mathbb{R}^{\lceil T/\sigma \rceil \times d}$ after a one-dimensional convolutional kernel process is applied to all the sub-sequences, and $\lceil T/\sigma \rceil$ is then replaced by $S$.

To ensure the inclusion of a wide range of subsequence motion patterns, we employed a random masking strategy to create the masked portion. We used a 60% masking rate to increase the difficulty of the reconstruction task.

### B. Temporal Representation

The visible encoder architecture follows the vanilla Transformer Encoder architecture. This architecture consists of multiple identical layers, each containing two sub-layers: a multi-head self-attention layer and a fully connected feedforward network. Each sub-layer is surrounded by residual connections with layer normalisation. Finally, each sublayer's output passes through a residual connection and feeds into the next layer. The model then outputs the positional embedding of the visible sequence $P \in \mathbb{R}^{S \times d}$. We added relative positional embeddings for each position of the subsequence embedding. The input embedding can be denoted by $Z = Z + P$, and $Z_v$ is fed into the encoder $E_\theta$, with subscripts v for visible and m for mask. The self-attention mechanism can simultaneously weigh all positions in the sequence with attention. Assuming an $L_v$-layer Transformer Block, the output layer $E_\theta^{L_v} = \{h_1^{L_v}, h_2^{L_v}, ..., h_{S_v}^{L_v}\}$ provides a globally visible sequence position representation. By using a masked encoder network, visible regions can be represented while excluding the representation of masked locations, thus eliminating discrepancies between pre-training and fine-tuning.

The decoupled encoder module for representation learning in the masked encoder architecture replaces the vanilla Transformer's self-attention with cross-attention. This allows for the decoupling of the visible and masked inputs. The decoupling module only embeds predictions for the mask position, while the output layer is represented by $D_\phi^{L_m} = \{f_1^{L_m}, f_2^{L_m}, ..., f_{S_m}^{L_m}\}$ for the contextual representation of the mask position.

### C. Masked Representation Regression (MRR) Optimization

$E_\xi$ represents a target encoder with identical hyperparameters to $E_\theta$, but optimized for $\xi$. We utilize this twin network structure as a target encoder to produce aligned representations. The target and decoupling encoders can respectively generate distinct views of the masked subsequence representations. Therefore, the task of optimisation involves minimising the alignment error between the two encoders: the target representation $E_\xi^{L_v}(Z_m)$ and the predicted representation $D_\phi^{L_m}(\tilde{Z})$. To account for the presence of anomalies, we use the more stable Huber Loss as the loss function, which is given in the following equation:

$$L_{alight} = \begin{cases} \frac{1}{2}(E_\xi^{L_v}(Z_m) - D_\phi^{L_m}(\tilde{Z}))^2, & \text{if } |E_\xi^{L_v}(Z_m) - \hat{E_\xi^{L_v}}(Z_m)| \leq \delta \\ \delta(|E_\xi^{L_v}(Z_m) - D_\phi^{L_m}(\tilde{Z})| - \frac{1}{2}\delta), & \text{otherwise} \end{cases} \quad (1)$$

The threshold parameter $\delta$ is where we aim for the behaviour of Huber Loss to be closer to the Mean Square Error (MSE), which we have set to 2. To prevent the model from collapsing, we have chosen to perform a stochastic optimisation step in the same way as TimeMAE. This involves minimising only the parameters of the on-line encoder while updating the target encoder with the Momentum Moving Average (MMAE).

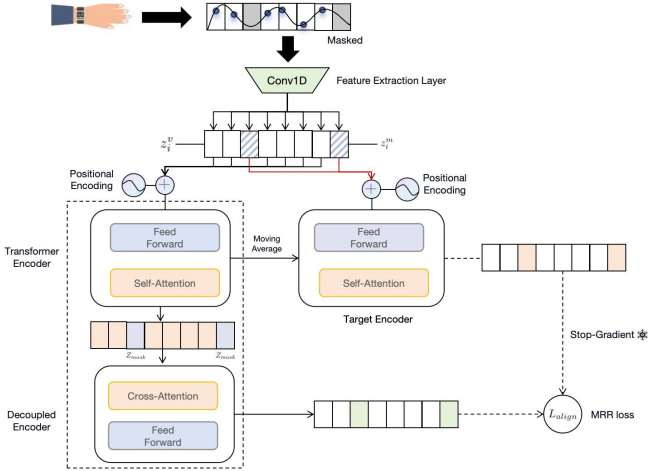

Figure 2. TimeMAE Extractor Module

### D. Attention Mechanism

After extracting temporal features using TimeMAE, we combine them with demographic features to create a one-dimensional feature vector $x = x_1, ..., x_i, ..., x_p \in \mathbb{R}^p$. However, since temporal data is inherently noisy and contains redundant temporal patterns, we introduce the Self-Reinforcing Attention Module (SRA) to allow the final classification layer to focus on more important features. This is in reference to TabSRA[4]. The SRA module encodes the input x as $K = [k_1, k_2, ..., k_i, ..., k_p]^T$ and $Q = [q_1, q_2, ..., q_i, ..., q_p]^T$. Where $k_i = (k_i^1, ..., k_i^{d_k}) \in \mathbb{R}^{d_k}$, $q_i = (q_i^1, ..., q_i^{d_k}) \in \mathbb{R}^{d_k}$ query matrices (Q) and keys (K) are generated by two independent fully connected feedforward networks. This embedding approach is effective in dealing with heterogeneous tabular data when there is an apparent feature interaction between the data. The use of this method not only reduces the need for multiple attention blocks but also significantly improves the efficiency of data processing. The keys in K are compared with the query Q on a case-by-case basis to quantify the alignment of different transformations for the same input. The attention weight $a = (a_1, ..., a_i, ..., a_p)$ is then calculated.

$$a_i = f(x) = \frac{q_i \times k_i}{d_k} \quad \text{for } i \in 1, ..., p \quad (2)$$

Finally, the attention score $a_i \in [0, 1]$ is obtained by scaling $d_k$. This score is then multiplied with the feature vectors $o = a \odot x$ to generate the output augmentation vectors $o = (o_1, ..., o_i, ..., o_p)$. These vectors are then fed into the linear layer for training, providing interpretability for subsequent work.

## III. EXPERIMENTS AND EVALUATION

### A. Experimental Design and Evaluation Indicators

The data sources used in this study were based on NHATS, a publicly available study. A total of 747 individuals (86%) provided Actidata. To generate minute-level physical activity (PA) intensity information, we combined the raw triaxial acceleration information for each timestamp into one-dimensional information by summing squares. For our analyses, we only considered acceleration data worn by participants during non-weekend periods on three consecutive days and excluded activity data between 11 p.m. and 5 a.m.. Thresholds were utilised for the composite scores of the SPPB. A score of 9 or less SPPB score indicated limited fitness, while a score of 9 or more indicated good fitness. Evaluation metrics such as AUC, Acc, F1_score, Recall, and Precision were employed, and the formula was calculated as follows:

$$\text{Accuracy} = \frac{\text{TP} + \text{TN}}{\text{TP} + \text{TN} + \text{FP} + \text{FN}} \quad (3)$$

$$\text{Recall} = \frac{\text{TP}}{\text{TP} + \text{FN}} \quad (4)$$

$$\text{Precision} = \frac{\text{TP}}{\text{TP} + \text{FP}} \quad (5)$$

$$\text{F1 Score} = \frac{2 \times \text{Precision} \times \text{Recall}}{\text{Precision} + \text{Recall}} \quad (6)$$

Where TP stands for True Positive, TN stands for True Negative, FP stands for False Positive, and FN stands for False Negative. AUC (Area Under the Curve) represents the area under the ROC curve, which is used to evaluate the performance of binary classification models.

The PyTorch deep learning framework was used to implement the network on a Windows platform with an 11th Gen Intel(R) Core(TM) i5-1135G7 @ 2.40GHz 2.42 GHz, 16G RAM, and an Intel(R) Iris(R) Xe Graphics Family GPU.

### B. Experimental Results

#### 1) Feature Ablation Experiments

To confirm the contribution of Actidata and demographic features to the model's performance, we conducted an ablation study on various hybrid features. The temporal features consist of 64-dimensional feature vectors extracted by TimeMAE-PFM. The classification layer utilises a combination of the SRA module and a linear layer.

The figure 3 displays the experimental results, indicating a significant improvement in the model's performance with the addition of mixed sequences. The AUC of the temporal features alone is 0.74, while the demographic features reach 0.79. However, with the inclusion of temporal features, the AUC reaches 0.84, resulting in an approximate 5% increase in accuracy and a more precise prediction.

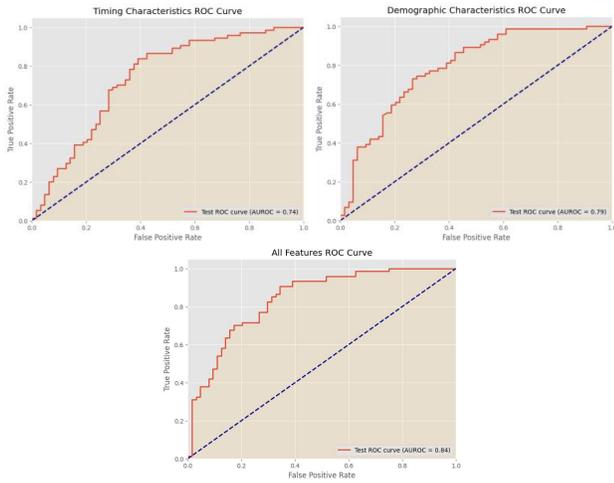

Figure 3. Comparison of results for different features

*2) Comparison of Indicators*

To confirm the validity and sophistication of our proposed model, we selected 11 cutting-edge timing algorithms[5-15], including TCN, XCM, and MiniRocket, for hybrid feature modelling. Each model was trained until loss convergence was achieved. The results shown in Figure 4 demonstrate that our model achieved 84.25% in AUC and 78.43% in F1, confirming its superiority. However, our model's accuracy is slightly lower than that of ResNet (75.18%) and MiniRocket (72.99%). Our structure consists solely of the SRA module with the linear classification layer, resulting in a significant reduction in the number of model parameters while maintaining optimal performance.

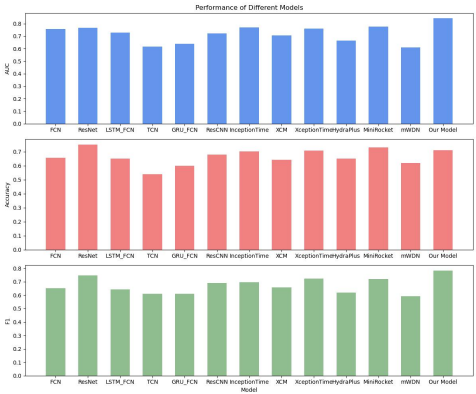

Figure 4. Advanced Timing Model Comparison

We attempted to replace the final classification layer with machine learning models including XGBoost, Random Forest, SVM, etc. We conducted five trainings and calculated the average and 95% confidence interval. Table 1 shows the final results, where we found that by simply adding the SRA module, the performance exceeded that of more complex machine learning models. Notably, our model, which is similar to logistic regression as a linear model, achieved an AUC of 0.822 (95% CI [0.798-0.842]), surpassing the AUC of 0.799 (95% CI [0.779-0.819]) of logistic regression. This demonstrates that the SRA module focuses on more important features, reduces interference caused by increasing feature dimensions, and showcases its superiority.

TABLE I. COMPARISON OF MACHINE LEARNING EFFECTIVENESS

| Model | AUROC | Accuracy | Sensitivity | Specificity | F1 Score | Precision | AUPRC |
|---|---|---|---|---|---|---|---|
| XGBoost | 0.75 [0.727-0.774] | 0.654 [0.636-0.671] | 0.926 [0.914-0.938] | 0.343 [0.314-0.373] | 0.74 [0.727-0.752] | 0.616 [0.603-0.629] | 0.771 [0.743-0.798] |
| Random Forest | 0.701 [0.675-0.726] | 0.654 [0.633-0.674] | 0.616 [0.588-0.644] | 0.696 [0.664-0.728] | 0.654 [0.635-0.673] | 0.698 [0.678-0.719] | 0.711 [0.678-0.745] |
| SVM | 0.758 [0.737-0.779] | 0.694 [0.673-0.715] | 0.706 [0.687-0.725] | 0.682 [0.644-0.72] | 0.711 [0.69-0.731] | 0.716 [0.684-0.749] | 0.767 [0.753-0.781] |
| Logistic | 0.799 [0.779-0.819] | 0.722 [0.707-0.737] | 0.743 [0.716-0.771] | 0.697 [0.686-0.707] | 0.739 [0.719-0.759] | 0.735 [0.719-0.752] | 0.804 [0.78-0.829] |
| Plain Bayesian | 0.674 [0.614-0.735] | 0.613 [0.557-0.669] | 0.703 [0.653-0.752] | 0.51 [0.411-0.608] | 0.659 [0.615-0.704] | 0.623 [0.573-0.673] | 0.698 [0.653-0.744] |
| Our Model | 0.822[0.798-0.842] | 0.706[0.662-0.756] | 0.699[0.565-0.884] | 0.733[0.548-0.840] | 0.720[0.664-0.784] | 0.763[0.695-0.807] | 0.825[0.801-0.853] |

*3) Cross-Domain Comparison*

The TimeMAE module is based on unsupervised learning, thus during the pre-training process, we consider using other datasets for pre-training. We selected the Human Activity Recognition (HAR) dataset[5], which also records demographic movements. The HAR dataset is used for human activity recognition and is collected from sensors, accelerometers, gyroscopes, and even devices like images and videos. The aim was to learn richer patterns of human kinetics from a larger dataset. The results show that features directly extracted after cross-domain adaptation, and processed through the classification layer of the SRA module, achieved an AUC of 0.810 [95% CI 0.802-0.819]. Furthermore, the average accuracy (0.710 [95% CI 0.683-0.738]) exceeded the pre-training effects on the existing dataset, highlighting the superiority of unsupervised learning. This demonstrates the commonality of human activities and provides guidance for accurately achieving cross-domain tasks in training large sequential models in the future.

TABLE II. COMPARISON OF CROSS-DOMAIN RESULTS

| Data | AUROC | Accuracy | Sensitivity | Specificity | F1 Score | Precision | AUPRC |
|---|---|---|---|---|---|---|---|
| NHATS->NHATS | 0.822[0.798-0.842] | 0.706[0.662-0.756] | 0.699[0.565-0.884] | 0.733[0.548-0.840] | 0.720[0.664-0.784] | 0.763[0.695-0.807] | 0.825[0.801-0.853] |
| HAR->NHATS | 0.810 [0.802-0.819] | 0.710 [0.683-0.738] | 0.791 [0.694-0.887] | 0.656 [0.552-0.760] | 0.755 [0.730-0.780] | 0.734 [0.697-0.771] | 0.811 [0.781-0.841] |

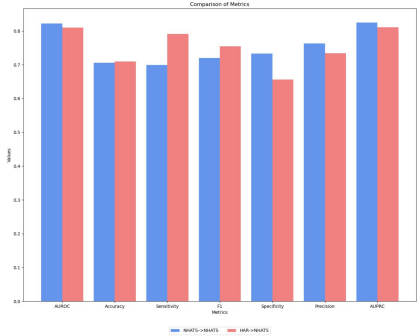

Figure 5.  Comparison of cross-domain effects

*4) Interpretability study*

The attention scores calculated by the SRA module were used to create a visualization and explain the model's decision-making process. The results are shown in Figure 6. It is evident that the demographic feature BMI is the most significant predictor of lower physical fitness as BMI increases, which aligns with reality. Additionally, the attentions obtained from the summation of temporal features are ranked second, fully reflecting the importance of temporal data for physical fitness monitoring. Taken together, our model offers a significant level of interpretability and is consistent with reality, providing guidance for physical fitness interventions.

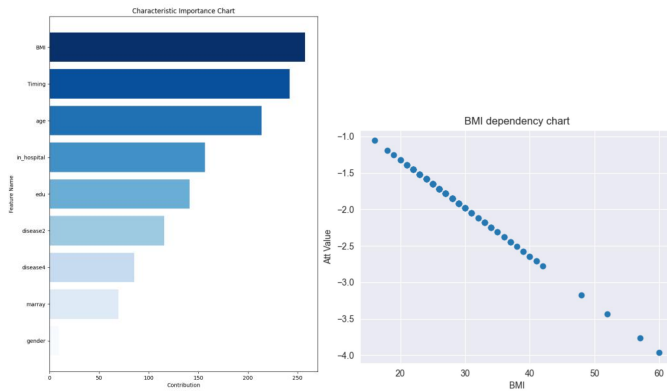

Figure 6.  Importance ranking (left) and BMI interaction (right)

## IV. CONCLUSIONS

This paper presents an innovative solution for PFM in the elderly that leverages Actidata and demographic information to enhance assessment accuracy. The incorporation of time self-supervised masked autoencoder pre-training effectively extracts temporal features. Additionally, a self-enhancing attention module enables deep fusion of multiple modalities. The method is validated using the NHATS dataset and compared to various methods under the same experimental setup. Results show that the method achieves an accuracy of 70.6% and an AUC of 82.20%. Future work will focus on refining the network model, exploring training on broader temporal datasets, and designing large-scale temporal pre-training models.

ACKNOWLEDGMENT

This work was supported by the National Key Research and Development Program of China (Grant No. 2023YFC3603800, 2023YFC3603802) and Sichuan Provincial student Innovation and Entrepreneurship Funding (C2024129211).

REFERENCES

[1] Guralnik J M, Simonsick E M, Ferrucci L, et al. A short physi cal performance battery assessing lower extremity function: association with self-reported disability and prediction of mortality and nursing home admission. Journal of gerontology, 1994, 49(2): M85-M94.

[2] DiPietro L. Physical activity in aging: changes in patterns and their relationship to health and function. The Journals of Gerontology Series A: Biological Sciences and Medical Sciences, 2001, 56(suppl 2): 13-22. doi.org/10.1093/gerona/56.suppl 2.13

[3] Cheng, M., Liu, Q., Liu, Z., Zhang, H., Zhang, R., & Chen, E. (2023). Timemae: Self-supervised representations of time series with decoupled masked autoencoders. arXiv preprint arXiv:2303.00320.

[4] Amekoe, K. M., Dilmi, M. D., Azzag, H., Dagdia, Z. C., Lebbah, M., & Jaffre, G. (2023, October). TabSRA: An Attention based Self-Explainable Model for Tabular Learning. In ESANN 2023-European Symposium on Artificial Neural Networks, Computational Intelligence and Machine Learning (pp. 199-204). Ciaco-i6doc. com.

[5] Reyes-Ortiz,Jorge, Anguita,Davide, Ghio,Alessandro, Oneto,Luca, and Parra,Xavier. (2012). Human Activity Recognition Using Smartphones. UCI Machine Learning Repository. https://doi.org/10.24432/C54S4K.

[6] Z. Wang, W. Yan, and T. Oates, Time series classification from scratch with deep neural networks: A strong baseline, International Joint Conference on Neural Networks (IJCNN), 2017. doi: 10.1109/ijcnn.2017.7966039.

[7] F. Karim, S. Majumdar, H. Darabi, and S. Chen, LSTM Fully Convolutional Networks for Time Series Classification, IEEE Access, 2018, pp.1662–1669. doi: 10.1109/access.2017.2779939.

[8] S. Bai, J. Z. Kolter, and V. Koltun, An Empirical Evaluation of Generic Convolutional and Recurrent Networks for Sequence Modeling, arXiv: Learning, 2018. doi: 10.48550/arXiv.1803.01271.

[9] N. Elsayed, A. S, and M. Bayoumi, Deep Gated Recurrent and Convolutional Network Hybrid Model for Univariate Time Series Classification, International Journal of Advanced Computer Science and Applications, 2019. doi: 10.14569/ijacsa.2019.0100582.

[10] J. Wang, Z. Wang, J. Li, and J. Wu, Multilevel Wavelet Decomposition Network for Interpretable Time Series Analysis, Proceedings of the 24th ACM SIGKDD International Conference on Knowledge Discovery \& Data Mining. 2018. doi: 10.1145/3219819.3220060.

[11] H. Ismail Fawaz et al., InceptionTime: Finding AlexNet for Time Series Classification, Data Mining and Knowledge Discovery, 2020, pp. 1936–1962. doi: 10.1007/s10618-020-00710-y.

[12] X. Zou, Z. Wang, Q. Li, and W. Sheng, Integration of residual network and convolutional neural network along with various activation functions and global pooling for time series classification, Neurocomputing, 2019, pp. 39–45. doi: 10.1016/j.neucom.2019.08.023.

[13] E. Rahimian, S. Zabihi, S. F. Atashzar, A. Asif, and A. Mohammadi, XceptionTime: A Novel Deep Architecture based on Depthwise Separable Convolutions for Hand Gesture Classification. Cornell University - arXiv, 2019. doi: 10.48550/arXiv.1911.03803.

[14] Kevin Fauvel, Tao Lin, Véronique Masson, Élisa Fromont, Alexandre Termier. XCM: An Explainable Convolutional Neural Network for Multivariate Time Series Classification. Mathematics , 2021, p. 3137. doi: 10.3390/math9233137.

[15] C. W. Tan, A. Dempster, C. Bergmeir, and G. I. Webb, MultiRocket: multiple pooling operators and transformations for fast and effective time series classification, Data Mining and Knowledge Discovery, 2022, pp. 1623–1646. doi: 10.1007/s10618-022-00844-1.